# Analyzing Journal Category Assignment Using a Paper-level Classification System: Multidisciplinary Sciences Journals

Jiandong Zhang[1], Liying Yang[2] and *Zhesi Shen[3]

[1] zhangjiandong21@mails.ucas.ac.cnt
School of Economics and Management, University of Chinese Academy of Sciences, Beijing, China; National Science Library, Chinese Academy of Sciences, Beijing, China

[2] yangly@mail.las.ac.cn
National Science Library, Chinese Academy of Sciences, Beijing, China

[3] shenzhs@mail.las.ac.cn
National Science Library, Chinese Academy of Sciences, Beijing, China

**Abstract**
In the field of scientometrics, the subject classification system of academic journals holds great importance. Accurate identification and classification of "multidisciplinary" journals are crucial in revealing the scientific structure and evaluating journals. Based on data from the Web of Science database from 2016 to 2020, we calculated the disciplinary diversity of journals using the paper-level subject classification system, then conducted a systematic analysis of JCR multidisciplinary journals. Studies showed that most multidisciplinary journals have high disciplinary diversity, while non-multidisciplinary journals tend to have relatively lower diversity. Some multidisciplinary journals with low disciplinary diversities may misclassify disciplines. In addition, there are inconsistencies in the diversity of journal disciplines at different granularities. Our study also visually analyzed the four types of diversity distribution tendencies of multidisciplinary journals. Moreover, ten potential multidisciplinary journals were found in non-multidisciplinary categories.

**Introduction**

A disciplinary classification system is essential to scientometric research. Generally, researchers need to assign publications or journals to research areas for further analysis (Waltman & van Eck, 2012). Understanding the characteristics of different classification systems is fundamental to research management because it affects subject knowledge acquisition, the evaluation of research impact, and even the establishment of research departments (Szomszor et al., 2021). "Multidisciplinary Science" is one part of the classification system used by the WoS. Besides the large publication scale of multidisciplinary journals, journals that have huge impacts on the world, such as *Nature*, *Science*, and *PNAS*, belong to this category, so there is a need to analyze the category assignment of multidisciplinary journals.

Journal-level and paper-level classification systems are the two main types of systems in recent research, and the former is more prevalent and widely used (Milojević, 2020). In the journal-level classification system, the research area of publications is subject to the area of their journal (Carpenter & Narin, 1973; Narin, 1976; Small & Griffith, 1974). Systems in Web of Science (WoS) and Scopus are two popular journal-level classification systems. WoS system is based on experts and citation analysis (Leydesdorff & Rafols, 2009), while the Scopus system is based on the title and content of the journal, evaluated by in-house experts[1].

The accuracy and reasonability of the journal-level classification system have been the focus of many investigations. Wang and Waltman (2016) identified journals with questionable classifications in the Web of Science and Scopus according to direct citation relations of journals between different categories. Su et al. (2019) found that the accuracy of the Chinese Science Citation Database journal classification system was approximately 50%. Besides these, researchers have raised concerns about the journal-level classification system. Leydesdorff and Bornmann (2014) pointed out that WoS did not provide sufficient analytical clarity for bibliometric normalization in evaluation practices due to the indexer effects. Rafols and Leydesdorff (2010) raised concerns that content-based classification systems were not designed to analyze the structure of science in science policy studies and the sociology of science. Van Eck et al. (2012) investigated the different research focuses in medical research areas with a visualization methodology and then pointed out that more complex bibliometric indicators could limit the field differences but still fail to account for heterogeneity within the domain in citation practice. However, the studies mentioned above do not focus on multidisciplinary sciences.

---
[1] https://service.elsevier.com/app/answers/detail/a_id/12007/c/10546/supporthub/scopus/

Usually, multidisciplinary journals are defined by a classification system that could be used in research evaluation researches (Su, 2020; Waltman, van Eck, 2019) and interdisciplinary research (Glanzel,2021; Glanzel,2021; Zhang et al., 2016; Zhang et al., 2018). Some journals which publish papers in various research areas cannot simply be divided into a specific field, so they usually are divided into multidisciplinary sciences. In the WoS classification system, journals classified as "multidisciplinary" cover entire large discipline, and journals that cover many fields are classified as multidisciplinary sciences, such as *Nature, Science*, and *PNAS* (Milojević, 2020). Current research on multidisciplinary science journals focuses on further subdividing the published papers, primarily based on citation information and filed cognitive words.

**Table 1 Main studies in subdividing multidisciplinary science journals**

| Authors | Method | Data | Year |
| --- | --- | --- | --- |
| Small & Griffith; Small & Koenig; Small, Sweeney & Greenlee | Co-citations | | 1974, 1977, 1985 |
| Glanzel, Schubert & Czerwon | Reference analysis | SCI, SSCI | 1999 |
| Debruin & Moed | Cognitive words from corporate addresses | *Nature, Science* | 1993 |
| Hua et al. | Reference analysis and cognitive words from corporate addresses | *Science* | 2015 |
| Ding, Ahlgren, Yang | Citation and cited information | *Nature, Science, PNAS* | 2018 |
| Milojević | Reference analysis | Web of Science Core Collection (1900-2017) | 2020 |

Finally, current related studies pay little attention to the classification accuracy of multidisciplinary sciences journals. This study aims to address this gap by employing the disciplinary diversity indicator to measure journals in multidisciplinary sciences. Then, we found features of journals in multidisciplinary sciences with visualization analysis of the scientific graph and recognized misclassified journals in multidisciplinary sciences and potential journals that could be classified as multidisciplinary.

**Data**

The journal list in this paper was downloaded from the Journal Citation Reports (2020) website, and 12,360 journals in SCI and SSCI were retrieved. The journal-level classification system is the Web of Science journal classification system, and the paper-level classification system is the Citation Topics paper classification system in the InCites database. 9,756,549 papers from 2016 to 2020 were retrieved from InCites database. In the final dataset, there were 12,225 journals due to the inconsistency between JCR and InCites. Citations among papers used in network construction were acquired from the local WoS database of the National Science Library, China.

The Web of Science journal classification system comprises 254 categories, covering natural sciences, social sciences, and art & humanities. Journals in this system are assigned to one or more categories. There were 73 journals in multidisciplinary sciences and 12,152 journals in other fields in this study.

The Citation Topics papers classification system uses the Leiden community cluster algorithm. This system has three levels (Macro-level, Meso-level, and Micro-level); each paper is assigned to only one topic at every level. The 2021 version of the system has 10 Macro-level topics, 326 Meso-level topics, and 2,444 Micro-level topics.

**Method**

*Disciplinary diversity of journals*

The disciplinary diversity indicator utilized in this study, LCDiv, is derived from the species diversity indicator in Ecology (Leinster, Cobbold, 2012). Several previous studies have employed it in disciplinary diversity measurement with the method of field citation similarity (Zhang, Rousseau, Glänzel, 2015; Zhang, Sun, Jiang, et al., 2021; Chen, Song, Shu, et al.,2022; Chen, Qiu, Arsenault,2021). The LCDiv can be expressed as follows:

$$\left(\sum_{i=1}^{n} p_i \left(\sum_{j=1}^{n} S_{ij} p_j\right)^{q-1}\right)^{\frac{1}{1-q}}$$

Previous studies usually considered q = 2, leading to:

$$\left(\sum_{i=1}^{n} p_i \left(\sum_{j=1}^{n} S_{ij} p_j\right)\right)^{-1} = \frac{1}{\sum_{i,j=1}^{n} S_{ij} p_i p_j}$$

where $s_{ij}$ is the similarity between fields $i$ and $j$, $p_i$ is the share of papers in the field $i$.

Citation similarity is used in most of the previous studies, but if there is no citation relationship between two papers, this method is difficult to obtain the similarity between them. To overcome this limitation, Node2Vec, a graph embedding method, was utilized in this study. Node2Vec can generate the distribution representation of a paper based on a citation network. It can get the cosine similarity between nodes with no direct citation relationship (Grover, Leskovec, 2016). In previous research, Node2Vec was successfully applied to cluster journals and produced better results than citations (Shen, Chen, Yang, et al., 2019). So, we used Node2Vec to represent research fields and calculate their cosine similarity.

In this study, we built the disciplinary citation network, $G(V, E, W)$. V is the collection of nodes of disciplines in the Citation Topics system, E is the collection of edges of citation relationship between disciplines, and W is the collection of weights, the numbers of citations. After that, the Node2Vec module was used to generate the 64-dimensional disciplinary vector, $Vector_i$, for each research area.

$$Vector_i = [v_1, v_2, \dots, v_{64}]$$

Then, $s_{ij}$ is the cosine similarity between $Vector_i$ and $Vector_j$.

$$S_{ij} = cosine(Vector_i, Vector_j)$$

Because each paper was assigned to 3 level classifications: macro-level, meso-level, and micro-level, we constructed three subject citation networks: Gmacro, Gmeso, and Gmicro. Gmacro comprised 10 nodes and 55 pairs of relations, Gmeso included 326 nodes and 51149 pairs of links, and Gmicro comprised 2,444 nodes and 1,432,610 pairs of relations. Subsequently, we calculated three collections of disciplinary similarity: Smacro, Smeso, and Smicro. Finally, based on the disciplinary diversity indicator, LCDiv, we obtained three collections of discipline diversity Dmacro, Dmeso, and Dmicro for each paper.

*An example of the disciplinary diversity computing process*

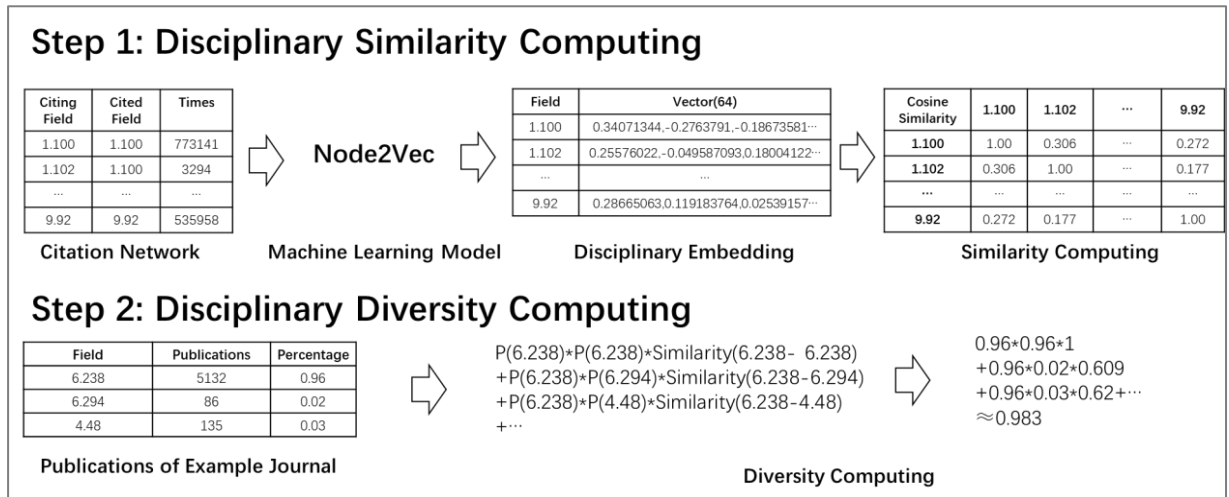

**Figure 1 The process used to compute diversity**

Figure 1 illustrates the process of computing disciplinary diversity for a sample journal at the meso-level. Firstly, disciplinary similarities are calculated using the Node2Vec model on the citation network from 2016 to 2020, generating a vector representation for each meso-field and enabling the calculation of pairwise cosine similarity scores. In the second step, the journal's publication output is counted, with the example journal publishing 5,132 papers in field 6.238, 86 papers in field 6.294, and 135 papers in field 4.48. Subsequently, the diversity of the journal's disciplines is calculated by combining the similarity scores with the proportion of publications in each field using the diversity formula.

*Visualization of scientific knowledge graph*

The scientific knowledge graph visualization process comprises three steps: network construction and layout computation, classification system mapping, and journal visualization.

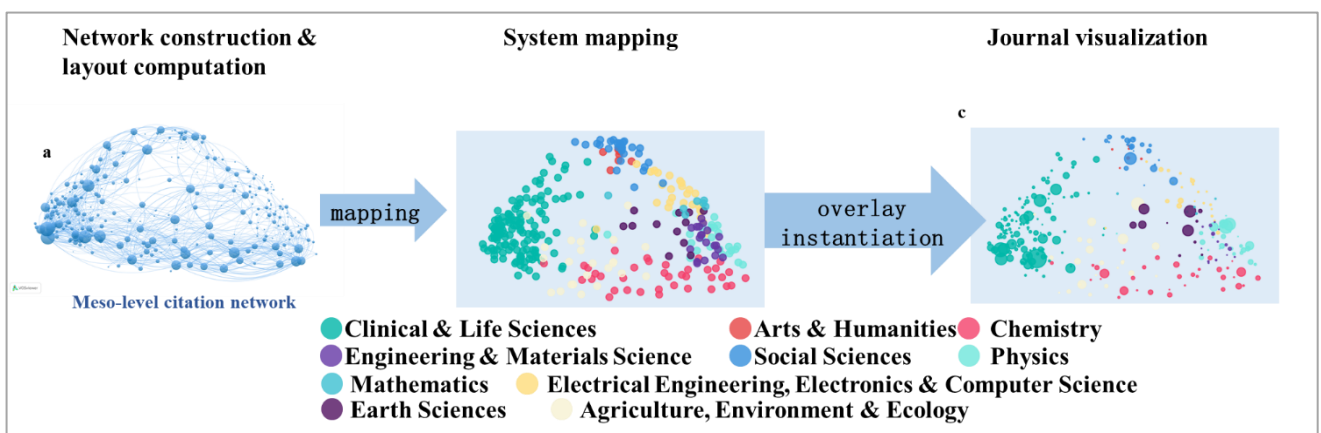

**Figure 2 Process of scientific knowledge graph visualization**

Network construction and layout computation: A citation network was built based on papers and citations between 2016 and 2020, then VOSviewer was used to generate the layout of this map. In Fig. 2a, each node represents a meso-level discipline, and each link denotes a pair of citation relationships. The nodes' size depends on their degrees; the citations between domains determine the link thickness. The larger the node, the more citations related to the field; the thicker the edge, the greater the number of citations between domains.

Classification system mapping: Each meso-level domain was mapped to its macro-level to obtain more information. So, we can judge the relationships of fields not only by their distances but also by their colors. In Fig.2b, each node represents a meso-level research area, with its color denoting the macro-level research area it belongs. There are 10 macro-level research areas. In the mapping process, areas that had only one paper were removed to reduce the classification system's noise.

Journal Visualization: In Fig. 2 takes *Nature* as an example of visualization. In this graph, each node represents the meso-level field of papers published between 2016 and 2020. Min-Max Normalization normalized the number of papers.

## Results

### Distribution of journal disciplinary diversity

Because journals of multidisciplinary sciences usually publish more papers from different filed than other journals, there may be differences between their distributions of journal disciplinary diversity.

**Table 2 Top 10 journals in three-level**

| Type | Journal | Diversity | | | Category |
|---|---|---|---|---|---|
| | | Micro | Meso | Macro | |
| Top 10 Micro-level Journals | Comptes Rendus De L Academie Bulgare des Sciences | 3.95 | 4.07 | 1.75 | Multidisciplinary Sciences |
| | Journal of Advanced Research | 3.93 | 4.12 | 1.59 | Multidisciplinary Sciences |
| | Scientific Reports | 3.82 | 3.91 | 1.44 | Multidisciplinary Sciences |
| | Interface Focus | 3.75 | 3.94 | 1.62 | Biology |
| | Sains Malaysiana | 3.73 | 3.81 | 1.58 | Multidisciplinary Sciences |
| | Scienceasia | 3.70 | 3.68 | 1.60 | Multidisciplinary Sciences |
| | Journal of King Saud University Science | 3.68 | 3.75 | 1.66 | Multidisciplinary Sciences |
| | Science | 3.68 | 3.75 | 1.59 | Multidisciplinary Sciences |
| | Nature | 3.64 | 3.71 | 1.57 | Multidisciplinary Sciences |
| | Chemie In Unserer Zeit | 3.62 | 3.71 | 1.46 | Chemistry, Multidisciplinary |
| Top 10 Meso-level Journals | Journal of Advanced Research | 3.93 | 4.12 | 1.59 | Multidisciplinary Sciences |
| | Comptes Rendus De L Academie Bulgare des Sciences | 3.95 | 4.07 | 1.75 | Multidisciplinary Sciences |
| | Interface Focus | 3.75 | 3.94 | 1.62 | Biology |
| | Scientific Reports | 3.82 | 3.91 | 1.44 | Multidisciplinary Sciences |
| | Sains Malaysiana | 3.73 | 3.81 | 1.58 | Multidisciplinary Sciences |
| | Science | 3.68 | 3.75 | 1.59 | Multidisciplinary Sciences |
| | Science Advances | 3.58 | 3.75 | 1.66 | Multidisciplinary Sciences |
| | Journal of King Saud University Science | 3.68 | 3.75 | 1.66 | Multidisciplinary Sciences |
| | Chemie In Unserer Zeit | 3.62 | 3.71 | 1.46 | Chemistry, Multidisciplinary |
| | Nature | 3.64 | 3.71 | 1.57 | Multidisciplinary Sciences |
| Top 10 Macro-level Journals | Siam Journal on Applied Mathematics | 2.75 | 2.86 | 1.95 | Mathematics, Applied |
| | Russian Journal of Numerical Analysis and Mathematical Modelling | 2.37 | 2.53 | 1.92 | Mathematics, Applied |
| | Entropy | 2.53 | 2.61 | 1.91 | Physics, Multidisciplinary |
| | Inverse Problems | 1.91 | 2.23 | 1.91 | Mathematics, Applied; Physics, Mathematical |
| | Open Physics | 2.51 | 2.63 | 1.89 | Physics, Multidisciplinary |

| | | | | |
|---|---|---|---|---|
| European Journal of Applied Mathematics | 2.43 | 2.46 | 1.89 | Mathematics, Applied |
| Fluctuation and Noise Letters | 2.23 | 2.24 | 1.89 | Physics, Applied; Mathematics, Interdisciplinary Applications |
| Physica D-Nonlinear Phenomena | 2.08 | 2.20 | 1.89 | Mathematics, Applied; Physics, Multidisciplinary; Physics, Fluids & Plasmas; Physics, Mathematical |
| Anziam Journal | 2.27 | 2.39 | 1.89 | Mathematics, Applied |
| Ima Journal of Applied Mathematics | 2.34 | 2.48 | 1.88 | Mathematics, Applied |

Table 2 shows the top 10 journals with the highest diversity at the micro-level, meso-level, and macro-level. At the micro-level and meso-level, most journals on the list belong to multidisciplinary sciences. Notably, prestigious journals such as Nature and Science are included. However, the macro-level results show a different trend from the micro-level and meso-level, with none of the journals belonging to multidisciplinary sciences.

Then, to explore the distribution characteristics in these three levels, we used a violin chart and statistical test (two-sided Mann-Whitney-Wilcoxon test) to analyze the whole journal collection.

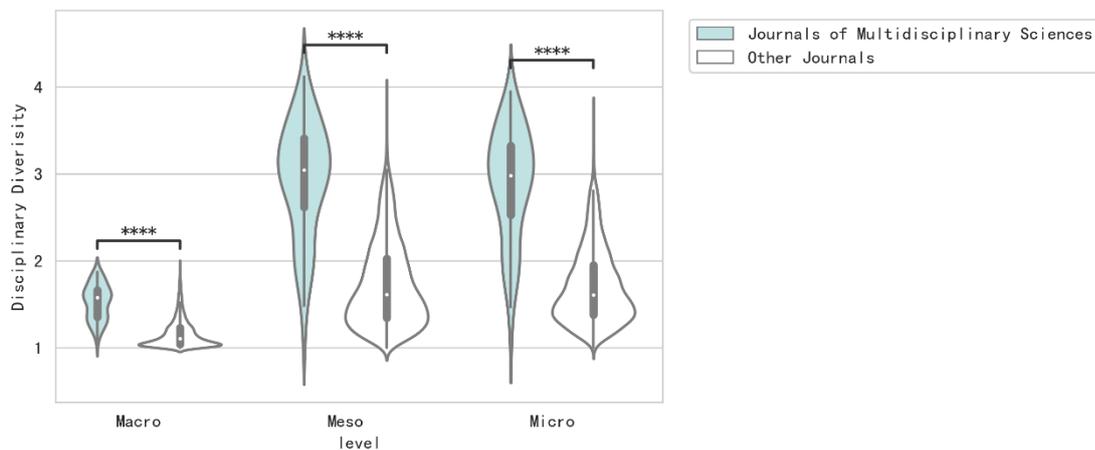

**Figure 3 Distributions of journal disciplinary diversity**

Distributions of journal disciplinary diversity divided into multidisciplinary science and other are shown in Fig.3. At the macro-level, the diversity of journals is approximately between 1 and 2. A significant difference exists between multidisciplinary sciences journals and others (P value is 1.941e-35). The diversity of non-multidisciplinary sciences journals is relatively concentrated, about 1.1, while the multidisciplinary sciences journals are more scattered, with an average of about 1.6. Similarly, at the meso-level, there is also a significant difference (P value is 6.005e-35), with multidisciplinary sciences journal diversity ranging from 1 to 4.5, and concentration approximately 3.1. Non-multidisciplinary journals are between 1 and 3.2, with a concentration of about 1.4. Similar distribution can be found at the micro-level and meso-level, and there is a significant difference in multidisciplinary sciences journals and others, too (P value is 6.119e-36).

We can further conclude that:

    1. A significant difference exists between the diversity distributions of multidisciplinary sciences journals and other journals. Disciplinary diversities of multidisciplinary sciences journals tend to concentrate on the upper part of the graph, while those of other journals concentrate on the lower value.

    2. Disciplinary diversities of non-multidisciplinary sciences journals tend to be more centralized, and multidisciplinary sciences journals tend to be more decentralized.

    3. some non-multidisciplinary sciences journals have high disciplinary diversity.

*Consistency of discipline diversity*

The previous section provided an overview of the data set's characteristics across three levels. This section aims to investigate the consistency of disciplinary diversity within the same journal across different levels. For example, whether a journal with a high value of macro-level diversity will have a high score at meso-level or micro-level.

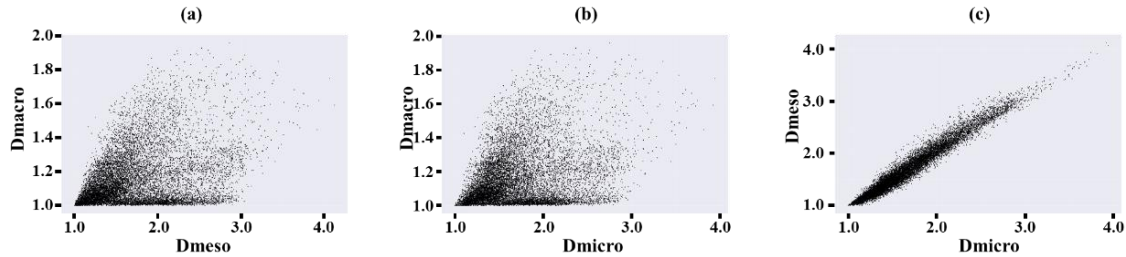

**Figure 4 Disciplinary diversity of each journal at different levels**

At first, we visualized the disciplinary diversity of each journal in a scatter plot. In Fig.4, each node represents a journal, and its location indicates its disciplinary diversity at three different levels. Taking Fig.4 (a) as an example. The x-axis represents the meso-level disciplinary diversity (Dmeso), while the y-axis denotes the macro-level disciplinary diversity (Dmacro). We can notice in Fig.3 that Fig.4 (a) and Fig.4(b) are more dispersed than Fig.4(c). By sorting the diversity of disciplines at each granularity and counting the overlap of the top 1000 journals at different levels, we found that the top 1000 journals at the macro-level diversity overlapped with 175 journals at the meso-level, and the micro-level overlapped 156 journals, while the meso-level and micro-level overlapped 884 journals.

**Table 3 Spearman correlation coefficient of disciplinary diversity at 3 levels**

| All categories journals/Multidisciplinary journals | Micro-level | Meso-level | Macro-level |
|---|---|---|---|
| Micro-level | - | 0.97***/0.98*** | 0.26***/0.27* |
| Meso-level | 0.97***/0.98*** | - | 0.30***/0.32** |
| Macro-level | 0.26***/0.27* | 0.30***/0.32** | - |

Then Spearman correlation coefficients of disciplinary diversity at 3 levels were calculated. The result reveals a strong correlation between micro-level and meso-level diversity, a weaker correlation between micro-level and macro-level diversity, and a weaker correlation between meso-level and macro-level diversity. Therefore, a journal may have quite different performances at different levels.

For example, the journal *Inverse Problems* has a macro-level disciplinary diversity value of 1.91 (ranked in the top 4/12225), a meso-level disciplinary diversity value of 2.23 (ranked 2019/12225), and a micro-level disciplinary diversity value of 1.91 (ranked 3403/12225). It indicates the journal is richer in macro-level disciplines than meso-level and micro-level subjects. In this journal, the top three macro-discipline fields with the highest proportion are 8 Earth Sciences (32.8%), 9 Mathematics (24.6%), and 1 Clinical & Life Sciences (13.0%). The three highest proportions of meso-disciplines are 8.212 Sensors & Tomography (30.8%), 9.162 Numerical Methods (15.8%), 1.175 Medical Physics (9.0%), and the three highest proportions of micro-disciplines are 8.212.1276 Microwave Imaging (23.7%), 9.162.1492 Inverse Heat Conduction Problem (12.8%), 1.175.444 Attenuation Correction (7.7%).

Among all journals, *Nature Reviews Disease Primers* has the most significant difference between the macro-level disciplinary diversity and meso-level rankings, with a difference of 11,795. Its macro-level disciplinary diversity is 1.00 (ranked 12068/12225), while its meso-disciplinary diversity and micro-disciplinary diversity are 2.90 (ranked 273/12225) and 2.80 (rated 255/12225), respectively. The journal with the most significant difference between disciplinary diversity at the meso-level and the micro-level is *Indogermanische Forschungen*, which ranks 4170. The maso-level subject diversity was 1.10 (ranked 6326/12225), and the values at the meso-level and micro levels were 1.65 (ranked 5780/12225) and 1.33 (rated 9950/12225).

*Composition of research topics in journals of multidisciplinary sciences*

In the previous section, we observed inconsistencies in macro-level, micro-level, and meso-level diversity. In this section, to explore the underlying reasons for these inconsistencies, we visualized the research topics of multidisciplinary sciences journals using a scientific graph that incorporated disciplinary diversity at the meso-level and macro-level. For example, why do some journals with high macro-level disciplinary diversity have middle or low meso-level diversity? We just analyzed the graph based on meso-level and macro-level dimensions for two reasons:
1. The meso-level classification comprises over 300 categories, and the micro-level classification contains over 2000 categories, making it difficult to analyze the visualization results based on these two dimensions.
2. Because of the strong correlation between micro-level and meso level diversity, the graph generated by micro-level and macro-level diversity would yield similar results to that generated by meso-level and macro-level diversity.

Multidisciplinary sciences journals were marked as red triangles in Fig.5a and others as gray dots. What can be seen is that the diversity of most multidisciplinary sciences journals is higher than that of other journals at the upper-right of the figure. But there are also some odd points, such as some multidisciplinary sciences journals with low macro-level and meso-level disciplines diversity, or some get a high meso-level diversity but low macro-level diversity.

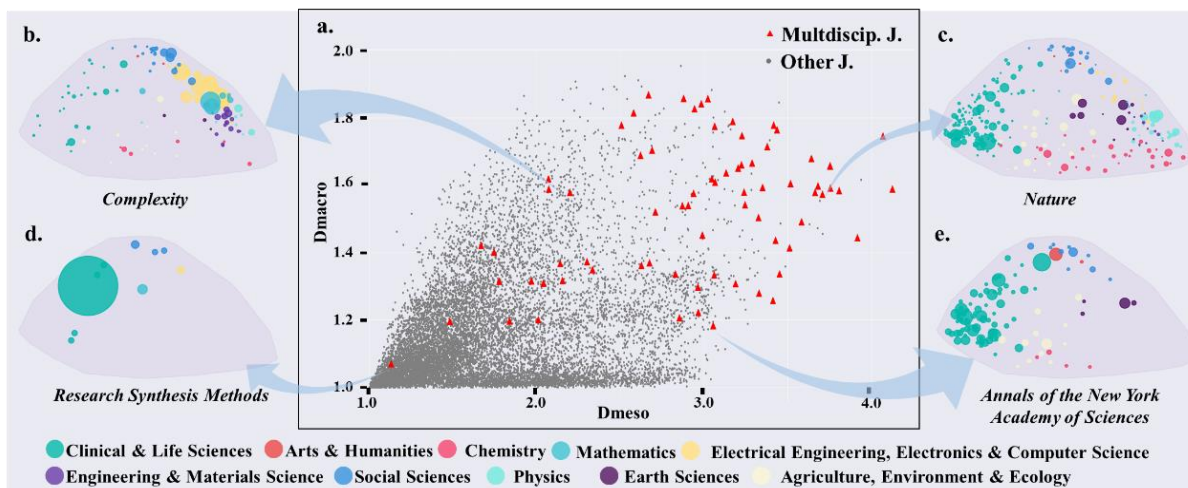

**Figure 5 Visualization examples of journals' topics**

Multidisciplinary sciences journals can be categorized into four types based on their degree of disciplinary diversities: high macro-level and meso-level diversity, high macro-level but low meso-level diversity, low macro-level but high meso-level diversity and low macro-level and meso-level diversity. After further analyzing the scientific graphs of these four kinds of journals, topics distribution characteristics of them can be found:

Journals with high disciplinary diversity at both macro- and meso-level have broad coverage of subjects at all levels. *Nature* is a typical example of such a journal. It can be seen from Fig. 5c that research papers published in *Nature* from 2016 to 2020 spanned multiple fields, including Life Sciences, Chemistry, Agriculture, Environment & Ecology, Physics, Arts and Humanities, and many other fields. Hence, there are many different colors in the graph. And these papers focus on many different topics in each macro research area, and the nodes are scattered even in the same color.

Journals with high disciplinary diversity at macro-level but low at meso-level cover a wide range of macro disciplines but fewer fields at meso-level. Unlike *Nature*, these journals show an interdisciplinary tendency, aiming to address relatively focused research problems using knowledge from multiple broad fields. For example, the journal *Complexity* mainly focuses on Electrical engineering, electronics & computer science, Social science, Engineering & materials, and other areas. Although it involves multiple macro-level disciplines, it centers on relatively concentrated scientific issues.

Journals with low disciplinary diversity at macro-level but high at meso-level publish fewer macro-level research areas but more meso-level research areas. These journals are similar to journals of professional multidisciplinary. For instance, the journal *Annals of the New York Academy of Science* published papers in some macro research areas such as Clinical & life sciences, Agriculture, environment & ecology. However, it focuses on research in the subarea of Clinical & life sciences.

Journals with low scores at macro-level and meso-level disciplinary diversity publish papers focusing on narrow areas. *Research Synthesis Methods* is an example of them. Papers in this journal primarily focus on problems in Evidence-Based Medicine. According to this feature, journals in the lower-left area may be misclassified in Multidisciplinary sciences.

*Recognition of potential journals in Multidisciplinary Sciences*

Some non-multidisciplinary journals are located in the upper right corner of Fig.6, and journals in this area have high macro-level and meso-level disciplinary diversity. These journals are potential multidisciplinary sciences journals.

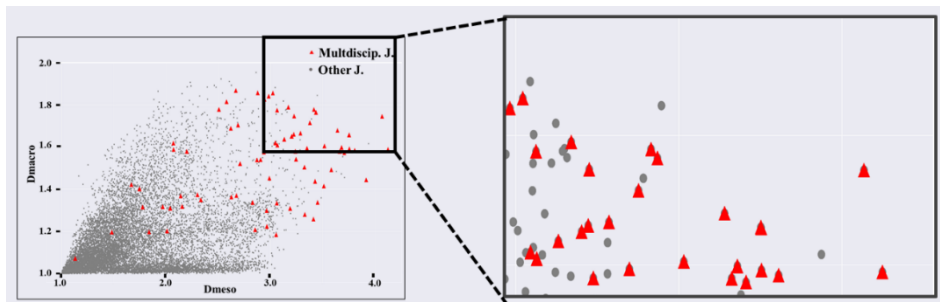

**Figure 6 Partial view of all journals**

To identify potential journals in Multidisciplinary Sciences, we used a distance-based method:
1. Normalizing the values of macro-level and meso-level diversity and defining them as the location of each journal. In this step, we used the max-min normalization;
2. Calculating the Euclidean distance between each point and point (1, 1). Point (1,1) is the highest one of the normalized value of macro-level and meso-level disciplinary diversity;
3. Finding the non-multidisciplinary sciences journals with the shortest distance.

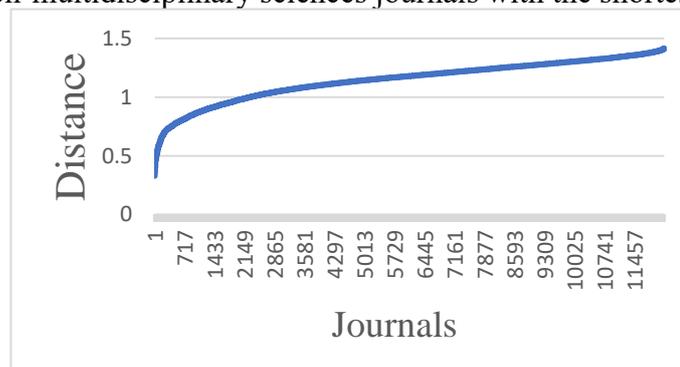

**Figure 7 Distances between non-multidisciplinary sciences journals**

Fig. 7 shows the distances between non-multidisciplinary sciences journal. Most distances of these journals are higher than 0.6. The distance increases fastest between 0 and 0.7; when the value exceeds 1, the value rises slowly.

**Table 4 Ten Journals with the Shortest Distances**

| Journal | Description on the official website | Distance |
|---|---|---|
| Engineering | The contents of our journal are based on the disciplines covered by the nine CAE divisions: • Mechanical and Vehicle Engineering• Information and Electronic Engineering• Chemical, Metallurgical, and Materials Engineering• Energy and Mining Engineering• Civil, Hydraulic, and Architecture Engineering• Agriculture• Environment&Light and Textile Industries Engineering• Medical and Health Care• Engineering Management | 0.24 |
| Micron | Micron is an interdisciplinary forum for all work that involves new applications of microscopy or where advanced microscopy plays a central role. The journal publishes research papers on the design, methods, application, practice or theory of microscopy and microanalysis, including reports on light optical, electron-beam, X-ray microtomography, ion microscopy and scanning-probe imaging. | 0.33 |
| Applied Sciences-Basel | Applied Sciences (ISSN 2076-3417) provides an advanced forum on **all aspects of applied natural sciences**. | 0.35 |
| SoftwareX | SoftwareX aims to acknowledge **the impact of software on today's research practice, and on new scientific discoveries in almost all research domains**. SoftwareX also aims to stress the importance of the software developers who are, in part, responsible for this impact. | 0.35 |
| Interface Focus | Each Interface Focus themed issue is devoted to a particular subject at the interface of the **physical and life sciences**. Formed of high-quality articles, they aim to **facilitate cross-disciplinary research across this traditional divide by acting as a forum accessible to all**. | 0.36 |
| Journal of Scientific & Industrial Research | This oldest journal of NISCAIR (started in1942) carries comprehensive reviews in **different fields of science & technology (S&T)**, including industry, original articles, short communications and case studies, on various facets of industrial development, industrial research, technology management, technology forecasting, instrumentation and analytical techniques | 0.36 |
| Electronics and Communications in Japan | ECJ aims to provide world-class researches in highly diverse and sophisticated areas of **Electrical and Electronic Engineering** as well as in related disciplines with emphasis on electronic circuits, controls and communications. | 0.37 |
| Hyle | HYLE is a double-blind peer-refereed international journal dedicated to **all philosophical aspects of chemistry**. | 0.37 |
| Frontiers in Physics | Frontiers in Physics publishes rigorously peer-reviewed research across the **entire field, from experimental, to computational and theoretical physics**. | 0.38 |
| Journal of Research of The National Institute of Standards and Technology | The Journal of Research of NIST reports NIST research and development in metrology and related fields of **physical science, engineering, applied mathematics, statistics, biotechnology, information technology**. | 0.38 |

Ten journals with the shortest distance are shown in Table 4. We also found the descriptions of each journal on their official website.

**Discussion and conclusion**

The precision and validity of the classification system have been a subject of great interest to researchers. Researchers have employed citation networks and case analysis methods to analyze the subject classification system at the journal and paper levels.

This study attempts to explore the characteristics of multidisciplinary journals from the perspective of the disciplinary diversity of papers included in multidisciplinary journals, aiming to provide new insights for journal classification research. Based on the method of Zhang et al. (2016), we proposed a method to measure the subject diversity of journals based on the paper's granular subject classification system. We compared the disciplinary diversities of journals at different levels. Diversities are calculated based on the paper's granularity of the subject classification system. We also visually analyzed the subject collection characteristics of different multidisciplinary sciences journals and identified some potential non-multidisciplinary journals.

Our research found that although the disciplinary diversity of multidisciplinary journals is generally higher than that of non-multidisciplinary journals, there are also some non-multidisciplinary journals with higher disciplinary diversity values. In addition, for the same journal, the performance of its disciplinary diversity will be affected by the granularity of the classification system used to calculate the indicators. The results calculated based on data with different granularity will be inconsistent. Therefore, caution is required when assessing the disciplinary diversity of different journals. The reasonable granularity for journal discipline diversity calculation is a possible future research direction.

For multidisciplinary journals, there are four distribution tendencies in the diversity of disciplines: multidisciplinary, interdisciplinary, professional, and non-multidisciplinary journals. Some non-multidisciplinary journals show high macro-level and meso-level diversity, suggesting their potential to be assigned to multidisciplinary category.

**Acknowledgement** Thanks to Prof. Ronald Rousseau for his valuable comments on the revision of the final draft. We also thank the anonymous reviewers for their suggestions for revision during the review process.